\documentclass[12pt]{iopart}
\usepackage{iopams}
  \expandafter\let\csname equation*\endcsname\relax
  \expandafter\let\csname endequation*\endcsname\relax
\usepackage{amsmath}
\usepackage{amssymb}
\usepackage{bm,color}
\usepackage{graphicx,epsfig,color}

\begin{document}

\title[]{Instabilities in the optical response of a Semiconductor Quantum Dot -- Metal Nanoparticle Heterodimer: Self-Oscillations and Chaos}

\author{Bintoro S Nugroho$^{1,2}$, Alexander A Iskandar$^{3}$, Victor~A~Malyshev$^{1}$ and Jasper Knoester$^{1,*}$}
\address{$^{1}$Zernike Institute for Advanced Materials, University of Groningen, Nijenborgh 4, 9747 AG Groningen, The Netherlands}
\address{$^{2}$ Fakultas Matematika dan Ilmu Pengetahuan Alam, Universitas Tanjungpura, Jl. Jendral A. Yani, 78124 Pontianak, Indonesia}
\address{$^{3}$ Physics of Magnetism and Photonics Research Group, Institut Teknologi Bandung, Jl. Ganesa 10, 40132 Bandung, Indonesia}
\address{$^{*}$e-mail: j.knoester@rug.nl}


\date{\today}

\begin{abstract}
\\
We theoretically investigate the nonlinear optical response of a heterodimer comprising a semiconductor quantum dot strongly coupled to a metal nanoparticle. The quantum dot is considered as a three-level ladder system with ground, one-exciton, and bi-exction states. As compared to the case of a two-level quantum dot model, adding the third (bi-exciton) state produces fascinating effects in the optical response of the hybrid system. Specifically, we demonstrate that the system may exhibit picosecond and sub-picosecond self-oscillations and quasi-chaotic behaviour under {\it single}-frequency continuous wave excitation. An isolated semiconductor quantum dot does not show such features. The effects originate from competing one-exciton and bi-exciton transitions in the semiconductor quantum dot, triggered by the self-action of the quantum dot via the metal nanoparticle. The key parameter that governs the phenomena mentioned is the ratio of the self-action strength and the bi-exciton shift. The self-oscillation regime can be achieved in practice, in particular, in a  heterodimer comprised of a closely spaced ZnS/ZnSe core-shell quantum dot and a spherical silver nanoparticle. The results may have applications in nanodevices for generating trains of ultrashort optical pulses.
\\
\\
Keywords: quantum dots, metal nanoparticles, hybrid nanostructures, exciton-plasmon interaction, nonlinear optical response, optical instabilities
\end{abstract}



\maketitle


\section{Introduction}
\label{Introduction}

Optical nonlinear effects occur when the intensity of applied electromagnetic fields are high enough to modify the material's optical properties. The first experimental demonstration of a nonlinear optical effect was second harmonic generation~\cite{FrankenPRL1961}, shortly after the first report of a working laser~\cite{MaimanNature1960}. Various phenomena arising from optical nonlinearities, have been applied in spectroscopic techniques to study physical properties of matter~\cite{NieAM1993,TominagaJCP1996,OronPRL2002} or to develop active optical devices~\cite{NieAM1993,CotterScience1999,NoskovNJP2012}. Despite their wide range of applications, processes related to optical nonlinearities are inherently weak, typically a few orders of magnitude smaller than those known in the linear regime~\cite{Boyd2003}. However, the nonlinear optical response of materials can be drastically enhanced by making use of plasmon-assisted effects. Optically excited plasmons in a metal produce a strong near-zone electromagnetic field~\cite{MaierAM2001,NovotnyNP2011}. This field can significantly increase the applied optical field, thereby giving rise to the subsequent enhancement of nonlinear interactions~\cite{WangJPCB2006,GongJCP2006,FuAPL2012,KauranenNP2012}.

In the past decade, nonlinear optical properties of nanoheterostructures consisting of quantum emitters [in particular, semiconductor quantum dots (SQDs)] coupled to metal nanoparticles (MNPs) have attracted considerable attention. The interest arises from the fact that such nanocomposites, due to hybridization of emitter states and plasmons, demonstrate new phenomena, such as enhancement of the second harmonic generation~\cite{SinghNT2013}, nonlinear and double-peaked Fano resonances~\cite{ZhangPRL2006,ArtusoNL2008}, optical bistability~\cite{MalyshevPRB2011,NugrohoJCP2013,LiOE2012} and many other interesting effects~\cite{RidolfoPRL2010,ManjavacasNL2011,ZhangPRB2011,ArtusoPRB2013,CarrenoJAP2014}. All these features originate from  the strong coupling between the constituents (MNPs and emitters), which can be controlled via the geometrical and material parameters of the hybrid cluster. This makes SQD-MNP nanoheterostructures promising nanomaterials with large and tunable nonlinearities, required for nonlinear optical applications.

Among the more exotic phenomena that can also be observed in nonlinear optical systems are optical instabilities, such as self-oscillations and chaos. It has been predicted theoretically by Ikeda~\cite{Ikeda1OC1979,IkedaPRL1980} that a ring cavity filled  with a nonlinear dielectric medium may show periodic and chaotic outputs. This phenomenon, called the Ikeda instability, has been experimentally demonstrated by Gibbs et al.~\cite{GibbsPRL1981} through the observation of optical turbulence and periodic oscillations in the optical bistability of a lead-based lanthanum-doped zirconate titanate sample. In several recent theoretical studies of nanoheterostructures, optical instabilities also have been predicted~\cite{ArtusoPRB2013,SadeghiAPL2014,SadeghiNT2015,SadeghiPRA2015,SadeghiJNR2016}. In particular, it has been demonstrated that a heterotrimer SQD-MNP-SQD, subject to an external optical field, may exhibit sustained oscillations and chaotic behaviour of the response~\cite{ArtusoPRB2013}. In addition, the optical response of an SQD-MNP heterodimer interacting with two applied fields, one in the visible and the other one in the infrared domain, has been found to reveal a periodic oscillatory behavior of the output field~\cite{SadeghiAPL2014,SadeghiNT2015,SadeghiPRA2015,SadeghiJNR2016}.

In this paper, we examine theoretically the nonlinear optical response of an SQD strongly coupled to a MNP, where the SQD is modelled as a ladder-like three-level system (ground, one-exciton, and bi-exciton states). We show that, depending on system's parameters, the SQD-MNP hybrid may exhibit, in addition to bistability~\cite{NugrohoPRB2015}, self-oscillations and quasi-chaotic behavior under {\it single}-frequency continuous wave (CW) excitation. This goes beyond the findings for {\it double}-frequency excitation, mentioned above~\cite{SadeghiAPL2014,SadeghiNT2015,SadeghiPRA2015, SadeghiJNR2016}. The effects originate from competing one-exciton and bi-exciton SQD transitions in the presence of a nearby MNP. The parameter that governs this behaviour is the ratio of the exciton-plasmon coupling and the bi-exciton shift. The new features in the response emerge when the SQD-MNP interaction becomes on the order of or larger than the bi-exciton shift. Previously~\cite{NugrohoPRB2015}, we reported on the optical response of a hybrid system consisting of a molecular dimer and an MNP, which has a bi-exciton shift that is much larger than the dimer-MNP coupling, thus not giving rise to the instabilities found here. In the case of a three-level ladder-like SQD, these two parameters can be fine-tuned.

This paper is organized as follows. In the next section, we present the system's model and the formalism for its description. In Sec. III, we report the results of our numerical calculations of the optical response of this system for a set of parameters that is achievable in practice and discuss these. In Sec. IV we summarize.

\section{Model and theoretical background}
\label{Model and method}
\begin{figure*}[ht!]
\begin{center}
\includegraphics[width=0.9\columnwidth]{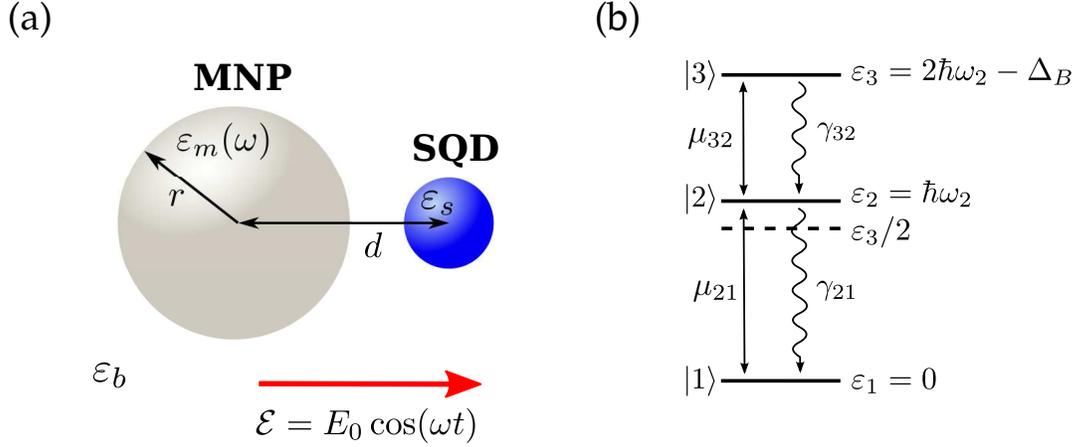}
\end{center}
\caption{(a)~Schematics of an SQD-MNP nanocomposite interacting with an applied field $\bm{\mathcal{E}}_0 = \boldsymbol{E}_0 \cos (\omega t)$. The field is linearly polarized along the system axis (shown by the red arrow). $d$ is the SQD-MNP center-to-center distance, $r$ is the radius of the MNP, $\varepsilon_s$ and $\varepsilon_m(\omega)$ are the dielectric constant of the SQD and the MNP, respectively. The system is embedded in an isotropic and non-absorbing medium with permittivity $\varepsilon_b$. (b)~Energy diagram of a ladder-type three-level SQD: $|1 \rangle$, $|2 \rangle$, and $|3 \rangle$ are the ground, one-exciton, and bi-exciton  states, respectively. The energies of corresponding states are $\varepsilon_1 =0$, $\varepsilon_2 =\hbar \omega_2$ and $\varepsilon_3 = 2\hbar\omega_2 -\Delta_B$. Here, $\Delta_B$ is the bi-exciton  binding energy. Allowed transitions with the corresponding transition dipole moments $\boldsymbol{\mu}_{21}$ and $\boldsymbol{\mu}_{32}$ are indicated by solid arrows. Wavy arrows denote the allowed spontaneous transitions with rates $\gamma_{32}$ and $\gamma_{21}$. The black-dashed horizontal line shows the location of the coherent two-photon resonance (with simultaneous absorption of two photons).}
\label{Fig1}
\end{figure*}

We theoretically investigate the nonlinear optical response of a nanocomposite consisting of an SQD and a nearby MNP embedded in an isotropic and non-absorbing background. This system is subjected to an applied CW optical field of frequency $\omega$ and amplitude $\boldsymbol{E}_0$, $\bm{\mathcal{E}}_0 = \boldsymbol{E}_0 \cos (\omega t)$, oriented along the system's axis [see Figure~\ref{Fig1}(a)]. The radii of the SQD ($a$) and the MNP ($r$), as well as their center-to-center distance ($d$) are assumed to be small compared to the optical wavelength, allowing us to use the quasi-static approximation for both particles~\cite{bohren2008,maier2007} and to neglect the retardation in the SQD-MNP interaction; the latter is dominated by the near-field multipolar coupling. For the MNP radii that we consider ($r > 5\mathrm{nm}$), size quantization effects are negligible~\cite{SchollNat2012}.
%
%

The optical excitation of an MNP is a localized surface plasmon, i.e. oscillations of free electrons. Like most previous studies of this type of nanohybrid, we adopt a classical description for the plasmons, using the MNP's polarizability $\alpha(\omega)$.
The latter is given by
\begin{equation}
  \alpha(\omega)= 4\pi r^3 \frac{\varepsilon_m(\omega)-\varepsilon_b}
                  {\varepsilon_m(\omega) + 2\varepsilon_b}~,
\label{alpha}
\end{equation}
where $\varepsilon_m$ is the dielectric function of the metal, and $\varepsilon_b$ is the permeability of the surrounding medium.

The optical excitations in the SQD are excitons. Generalizing previous studies~\cite{ZhangPRL2006,ArtusoNL2008,MalyshevPRB2011,LiOE2012,NugrohoJCP2013}, which were restricted to a two-level model of the SQD (a ground and a one-exciton state), here we incorporate also a bi-exciton state in the SQD, corresponding to two excitations coupled by the Coulomb interaction.  In such a system, the degeneracy of the one-exciton state is lifted due to the anisotropic electron-hole exchange, leading to two split linearly polarized one-exciton states (see, e.g., Refs.~\cite{StuflerPRB2006,JundtPRL2008,GerardotNJP2009}). In this case, the ground state is coupled to the bi-exciton state via the linearly polarized one-exciton transitions. By choosing the appropriate polarization of the applied field, i.e. selecting one of the single-exciton states, the system effectively acquires a three-level structure with a ground state $|1\rangle$, one exciton state $|2\rangle$, and bi-exciton  state $|3\rangle$ with corresponding energies $0$, $\hbar\omega_{2}$, and $\hbar\omega_3 = 2\hbar\omega_2 - \Delta_B$, where $\Delta_B$ is the bi-exciton shift. Within this model, the allowed transitions, induced by the external field, are $|1\rangle \leftrightarrow |2\rangle$ and $|2\rangle \leftrightarrow |3\rangle$, which are characterized by the transition dipole moments $\boldsymbol{\mu}_{21} (= \boldsymbol{\mu}_{12})$ and $\boldsymbol{\mu}_{32} (= \boldsymbol{\mu}_{23})$, respectively (for the sake of simplicity, we assume that they are real). The states $|3\rangle$ and $|2\rangle$ spontaneously decay with rates $\gamma_{32}$ and $\gamma_{21}$, accordingly [see Fig.~\ref{Fig1}(b)]. Note that the bi-exciton  state $|3\rangle$, having no allowed transition dipole moment from the ground state $|1\rangle$, can be reached either via consecutive $|1\rangle \rightarrow |2\rangle \rightarrow |3\rangle$ transitions or via the simultaneous absorption of two photons of frequency $\omega_3/2$.

The optical dynamics of the SQD is described by means of the Lindblad quantum master equation for the density operator $\rho(t)$, which in the rotating (with frequency $\omega$ of the external field) frame reads
\begin{subequations}
\label{MasterEqAndHamiltonian}
\begin{align}
\begin{split}
\dot{\rho} = -\frac{i}{\hbar} \left[{H^{RWA}},\rho\right ]
           + \frac{\gamma_{21}}{2} \left( \left[ \sigma_{12} \rho,
           \sigma_{21}\right]  + \left[ \sigma_{12},
           \rho\, \sigma_{21}\right]\right)
           + \frac{\gamma_{32}}{2} \left( \left[
           {\sigma}_{23}{\rho},\sigma_{32}\right]
           + \left[ \sigma_{23},\rho\, \sigma_{32}\right]\right)
\end{split}
\label{DensityMasterEq}\\
\begin{split}
H^{RWA} =  \hbar \left( \Delta_{21}\sigma_{22} + \Delta_{31}\sigma_{33} \right)
        -  \hbar \left( {\Omega}_{21} \sigma_{21} + \Omega_{32} \sigma_{32} \right) + h.c.
\end{split}
\label{HamiltonianRWA}
\end{align}
\end{subequations}
In Eq.~(\ref{DensityMasterEq}), $H^{RWA}$ is the SQD Hamiltonian in the rotating wave approximation, square brackets denote the commutator, and the two other terms represent the relaxation operator, where $\sigma_{ij} = |i\rangle \langle j|$ \, ($i,j = 1,2,3$). In Eq.~(\ref{HamiltonianRWA}), $ \Delta_{21} = \omega_2 - \omega$ and $\Delta_{31} = \omega_3 - 2\omega$ are, in fact, the energies of states $|2 \rangle$ and $|3 \rangle$ in the rotating frame, accordingly. Alternatively, these quantities may be interpreted as
the detunings away from the one-photon resonance and the coherent two-photon resonance, respectively. $\Omega_{21} = \boldsymbol{\mu}_{21} \boldsymbol{E}_{SQD}/(2\hbar)$ and $\Omega_{32} = \boldsymbol{\mu}_{32} \boldsymbol{E}_{SQD}/(2\hbar)$ are Rabi frequencies for the corresponding transitions, where $\boldsymbol{E}_{SQD}$ is the amplitude of the total field acting on the SQD. The latter is the sum of the applied field ${\boldsymbol{E}}_0$ and the field produced by the plasmon oscillations in the MNP~\cite{ArtusoNL2008,MalyshevPRB2011,NugrohoJCP2013}.
For the MNP's field, we take into account the contribution of higher multipoles, which is important if the MNP's radius $r$ is on the order of the SQD-MNP spacing $d$ (our case, see below). Then $\boldsymbol{E}_{SQD}$ reads~\cite{YanPRB2008,ArtusoPRB2011}
\begin{equation}
\boldsymbol{E}_{SQD} = \frac{1}{\varepsilon_s'}\left[ 1
                     +\frac{\alpha(\omega)}{2 \pi d^3} \right]\boldsymbol{E}_0
                     + \frac{
                     \boldsymbol{\mu}_{21}\rho_{21}+\boldsymbol{\mu}_{32}\rho_{32}}
                     {16 \pi^2 \varepsilon_0 \varepsilon_b \varepsilon_s'}
                     \sum_{n=1}^{\infty}\frac{(n+1)^2\alpha_n(\omega)}{d^{2n+4}}~,
\label{E_SQD}
\end{equation}
were $\varepsilon_s' = (\varepsilon_s + 2 \varepsilon_b)/(3\varepsilon_b)$ is the effective dielectric function of the SQD, $\varepsilon_s$ is the bulk dielectric function of the semiconductor, and
\begin{equation}
\label{alpha_n}
\alpha_n(\omega) = 4\pi r^{2n+1} \frac{\varepsilon_m(\omega)-\varepsilon_b}{\varepsilon_m(\omega) + \frac{n+1}{n}\varepsilon_b} \ .
\end{equation}
The first term in Eq.~(\ref{E_SQD}) describes the renormalization of the external field due to the presence of the MNP. The second one represents the electromagnetic self-action of the SQD via the MNP, in which the field acting on the SQD depends on the SQD state itself through the density matrix elements $\rho_{21}$ and $\rho_{32}$. As will be shown below, this plays an important role in the optical response of the SQD-MNP hybrid,

Using Eq.~(\ref{E_SQD}), we can specify expressions for the Rabi frequencies $\Omega_{21}$ and $\Omega_{32}$:
\begin{subequations}
\label{Rabi frequencies}
\begin{align}
\begin{split}
        \Omega_{21} = \widetilde{\Omega}^0_{21} + G_1 \rho_{21} + G_3 \rho_{32} \ ,
\end{split}
\label{Omega21}\\
\begin{split}
        \Omega_{32} = \widetilde{\Omega}^0_{32} + G_3 \rho_{21} + G_2 \rho_{32} \ ,
\end{split}
\label{Omega32}
\end{align}
\end{subequations}
where we introduced:
\begin{subequations}
\label{All_Omega}
\begin{align}
\begin{split}
        \widetilde{\Omega}^0_{21} = \frac{1}{\varepsilon_s'}\left[ 1
                     +\frac{\alpha(\omega)}{2 \pi d^3} \right]\Omega^0_{21} \ ,
\end{split}
\label{Omega021}\\
\begin{split}
        \widetilde{\Omega}^0_{32} = \frac{1}{\varepsilon_s'}\left[ 1
                     +\frac{\alpha(\omega)}{2 \pi d^3} \right]\Omega^0_{32} \ ,
\end{split}
\label{Omega032}
\end{align}
\end{subequations}
and
\begin{subequations}
\label{All_G}
\begin{align}
\begin{split}
        G_1 = \frac{\boldsymbol{\mu}_{21}\boldsymbol{\mu}_{21}}{16 \pi^2 \varepsilon_0 \varepsilon_b \varepsilon_s'}\sum_n \frac{(n+1)^2\alpha_n(\omega)}
                     {d^{2n+4}} \ ,
\end{split}
\label{G1}\\
\begin{split}
        G_2 = \frac{\boldsymbol{\mu}_{32}\boldsymbol{\mu}_{32}}
                     {16 \pi^2 \varepsilon_0 \varepsilon_b \varepsilon_s'}
                     \sum_n \frac{(n+1)^2\alpha_n(\omega)}
                     {d^{2n+4}}\ ,
\end{split}
\label{G2}\\
\begin{split}
        G_3 = \frac{\boldsymbol{\mu}_{21}\boldsymbol{\mu}_{32}}
                     {16 \pi^2 \varepsilon_0 \varepsilon_b \varepsilon_s'}
                     \sum_n \frac{(n+1)^2\alpha_n(\omega)}
                     {d^{2n+4}}\ .
\end{split}
\label{G3}
\end{align}
\end{subequations}
Here, $\Omega^0_{21} = \boldsymbol{\mu}_{21} \boldsymbol{E}_0/(2\hbar)$ and $\Omega^0_{32} = \boldsymbol{\mu}_{32} \boldsymbol{E}_0/(2\hbar)$ are the Rabi frequencies of the external field for the corresponding transitions. The complex value quantities $G_1 = G_1^R + iG_1^I$, $G_2 = G_2^R + iG_2^I$, and $G_3 = G_3^R + iG_3^I$ represent the so-called feedback parameters, describing the self-interaction of the SQD via the MNP~\cite{ArtusoNL2008,MalyshevPRB2011,NugrohoJCP2013}.

Using the above, the equations for the density matrix elements $\rho_{ij}$ \, ($i,j = 1,2,3$), describing the optical dynamics of the SQD in the presence of the MNP, read
\begin{subequations}
\label{all_dR1}
\begin{align}
\begin{split}
\dot{\rho}_{11} &= \gamma_{21} \rho_{22} + i(\widetilde{\Omega}^{0*}_{21} \rho_{21}
                - \widetilde{\Omega}^0_{21} \rho^*_{21}) + i (G^*_3 \rho^*_{32} \rho_{21}
                - G_3 \rho_{32} \rho^*_{21}) + 2 G_{1}^I \rho_{21} \rho^*_{21}~,
\end{split}
\label{rho11}\\
\begin{split}
\dot{\rho}_{22} &= - \gamma_{21} \rho_{22} + \gamma_{32} \rho_{33}
                +  i (\widetilde{\Omega}^0_{21} \rho^*_{21} - \widetilde{\Omega}^{0*}_{21} \rho_{21} +  \widetilde{\Omega}^{0*}_{32} \rho_{32}             - \widetilde{\Omega}^0_{32} \rho^*_{32} )\\
                &\quad + 2i G_{3}^R (\rho_{32} \rho^*_{21} - \rho^*_{32} \rho_{21})
                + 2 (G_{2}^I \rho_{32}  \rho^*_{32} - G_{1}^I \rho_{21} \rho^*_{21})~,
\end{split}
\label{rho22}\\
\begin{split}
\dot{\rho}_{33} &= -\gamma_{32} \rho_{33} + i (\widetilde{\Omega}^0_{32} \rho^*_{32}
                - \widetilde{\Omega}^{0*}_{32} \rho_{32})
               + i (G_3 \rho^*_{32} \rho_{21} - G^*_3 \rho_{32} \rho^*_{21})
                - 2 G_{2}^I \rho_{32} \rho^*_{32}~,
\end{split}
\label{rho33}\\
\begin{split}
\dot{\rho}_{21} &= -\left[ i \left(\Delta_{21} + G_{1}^R Z_{21} \right)
                +\frac{1}{2}\gamma_{21} - G_{1}^I Z_{21} \right] \rho_{21}\\
                &\quad + i ( \widetilde{\Omega}^{0*}_{32} \rho_{31}
                - \widetilde{\Omega}^0_{21} Z_{21} )
                + i \left[ (G^*_3 \rho^*_{21} + G^*_2 \rho^*_{32}) \rho_{31}
                - G_3 \rho_{32} Z_{21} \right]~,
\end{split}
\label{rho21}\\
\begin{split}
\dot{\rho}_{32} &= -\left[ i(\Delta_{32} + G_{2}^R Z_{32})
                + \frac{1}{2} (\gamma_{32} + \gamma_{21})
                -  G_{2}^I Z_{32} \right] \rho_{32}\\
                &\quad- i ( \widetilde{\Omega}^{0*}_{21} \rho_{31}
                + \widetilde{\Omega}^0_{32} Z_{32} )
                - i \left[ (G^*_1 \rho^*_{21} + G^*_3 \rho^*_{32}) \rho_{31}
                + G_3 \rho_{21} Z_{32} \right]~,
\end{split}
\label{rho32}\\
\begin{split}
\dot{\rho}_{31} &= -\left(i\Delta_{31} + \frac{1}{2}\gamma_{32} \right) \rho_{31}
                + i (\widetilde{\Omega}^0_{32} \rho_{21}
                - \widetilde{\Omega}^0_{21} \rho_{32}) + i (G_2-G_1) \rho_{32} \rho_{21}\\
                &\quad+  i G_3 (\rho_{21} \rho_{21} - \rho_{32} \rho_{32})~,
\end{split}
\label{rho31}
\end{align}
\end{subequations}
where $\Delta_{32} = \omega_3 - \omega_2 - \omega$ is the detuning away from the $|3 \rangle \leftrightarrow |2 \rangle$  and $Z_{ji} = \rho_{jj}-\rho_{ii}$ is the population difference between the states $|j\rangle$ and  $|i\rangle$.

As follows from Eqs.~(\ref{rho11}) -~(\ref{rho31}), the self-action of the SQD, governed by the feedback parameters $G_i$ \, ($i,j = 1,2,3$), gives rise to many additional nonlinearities as compared to an isolated SQD. The two of these that should be mentioned especially, are (i) - renormalization of the SQD frequencies, $\omega_2 \rightarrow \omega_2 + G_1^R Z_{21}$ and  $\omega_3 \rightarrow \omega_3 + G_2^R Z_{32}$, and (ii) - renormalization of the relaxation rates of the off-diagonal density matrix elements, $\gamma_{21}/2 \rightarrow \gamma_{21}/2 - G_1^I Z_{21}$ and $(\gamma_{21} + \gamma_{32})/2 \rightarrow (\gamma_{21} + \gamma_{32})/2 - G_2^I Z_{32}$ [see Eqs.~(\ref{rho21}) and~(\ref{rho32})], both depending on the corresponding population differences. As will be shown below, these two effects are essential in the formation and understanding of the complicated optical response of the hybrid system.


\section{Numerical results and discussion}
\label{Results and discussion}

In this section, we analyse the optical response of an SQD in close proximity to a spherical Ag MNP, both embedded in a host with  permittivity $\varepsilon_b = 2.16$ (silica). The Ag nanosphere considered has radius $a = 11~\mathrm{nm}$. The dielectric function $\varepsilon_m$ of silver is taken from experiment~\cite{JohnsonPRB1972}. The corresponding surface plasmon resonance is found to be  $\omega_{sp} = 3.12$ eV. For the SQD, we use the following set of parameters: the dielectric constant $\varepsilon_s = 6$, the population relaxation rates, describing the spontaneous emission of states $|2\rangle$ and $|3\rangle$, are $\gamma_{21} = 1/200~\mathrm{ps^{-1}}$ and  $\gamma_{32} = 1/300~\mathrm{ps^{-1}}$. The transition dipole moments are $\mu_{21}=0.60~e~\mathrm{nm}$ and $\mu_{32} = 0.75~e~\mathrm{nm}$, and the energies of the bare one-exciton and bi-exciton transitions are $\hbar \omega_2 = 3.1~\mathrm{eV}$ and $\hbar \omega_3 = \hbar (2\omega_2 - \Delta_B)$ with $\hbar\Delta_B = 2.5~\mathrm{meV}$, parameters that can be realized by tuning the geometry of ZnS/ZnSe core-shell quantum dots.~\cite{SenJPCS2010} The center-to-center SQD-MNP distance is taken to be $d = 16~\mathrm{nm}$.

The system is subjected to an external field under two resonance conditions: $\omega=\omega_{2}$ (in resonance with the bare one-exciton transition) and $\omega = \omega_3/2$ (in resonance with the coherent two-photon transition). We analyse the optical response of the hybrid by examining the populations $\rho_{11}$, $\rho_{22}$, $\rho_{33}$, and the magnitude of the SQD's dipole moment, $|p_{SQD}| = |\mu_{21}\rho_{21} + \mu_{32}\rho_{32}|$. The applied field magnitude $\Omega^0_{21}$ is swept up and down adiabatically in order to check whether the optical response depends on the history of the input.

\subsection{Isolated SQD}
\label{Isolated SQD}

\begin{figure}[ht!]
\begin{center}
\includegraphics[width=0.8\columnwidth]{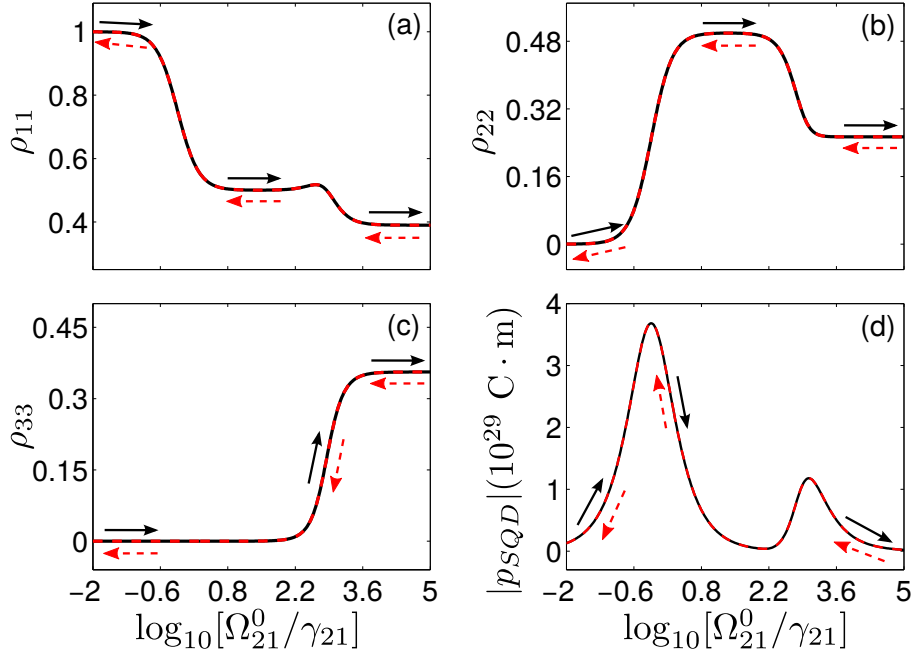}
\end{center}
\caption{Field dependent response of an isolated SQD for $\omega = \omega_2$. Populations of the ground state $\rho_{11}$ (a), one-exciton state $\rho_{22}$ (b), and bi-exciton   state $\rho_{33}$ (c). (d) SQD's dipole moment magnitude $|p_{SQD}|$. Solid and dashed arrows show the direction of sweeping the applied field magnitude $\Omega^0_{21}$ up and down, respectively. The field is in resonance with the one-exciton transition ($\omega = \omega_2$). System parameters are described in the text.
}
\label{Fig2}
\end{figure}

\begin{figure}[ht!]
\begin{center}
\includegraphics[width=0.8\columnwidth]{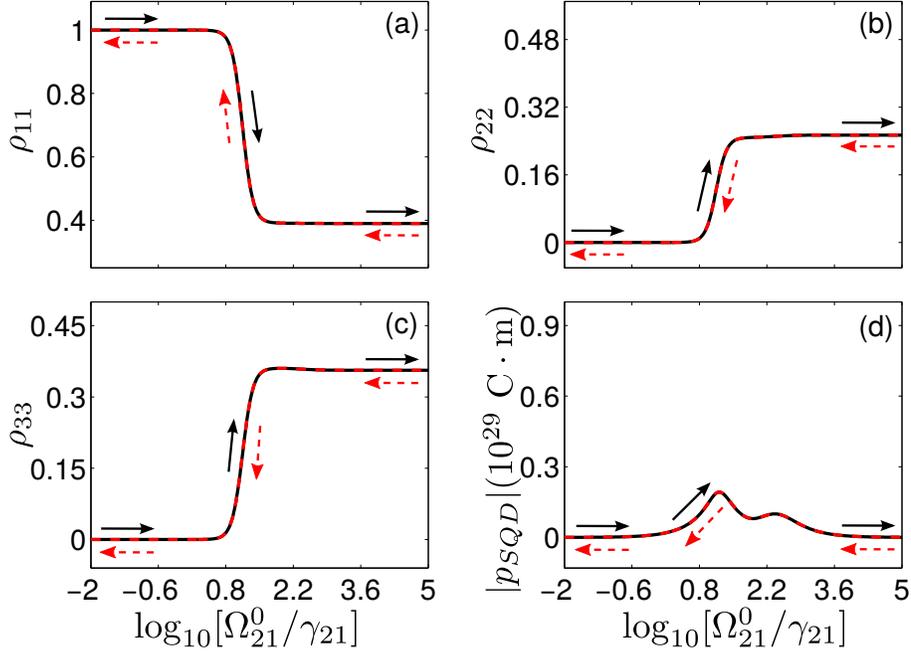}
\end{center}
\caption{Same as in Fig.~\ref{Fig2}, but now for the external field in resonance with the coherent two-photon transition ($\omega = \omega_3/2 = \omega_2 - \Delta_B/2$).
}
\label{Fig3}
\end{figure}

First, we investigate the optical response of a quasi-isolated SQD, by setting the SQD-MNP centre-to-centre distance to $d = 200$ nm, which is  large enough to decouple both particles. At such a distance, $|G_m| \ll \gamma_{21}, \gamma_{32}$.
We calculate the field dependent behaviour of $\rho_{11}$, $\rho_{22}$, $\rho_{33}$, and $|p_{SQD}| = |\mu_{21}\rho_{21} + \mu_{32}\rho_{32}|$ for two resonance conditions: (i) - the applied field is in resonance with the bare one-exciton transition $\omega=\omega_1$ [see Fig.~(\ref{Fig2})] and (ii) - it is tuned to the two-photon resonance, $\omega = \omega_3/2 = \omega_2 - \Delta_B/2$ [see Fig.(\ref{Fig3})]. As mentioned above, we adiabatically swept the applied field magnitude $\Omega^0_{21}$ up and down. The trajectories of increasing and decreasing $\Omega^0_{21}$ are indicated in Fig.~(\ref{Fig2}) and Fig.(\ref{Fig3}) by black-solid and red-dashed arrows, respectively.

In Fig.~\ref{Fig2}, the field dependent response of the SQD is shown for the case of excitation at the single exciton resonance, $\omega = \omega_2$. For a weak external field ($\Omega^0_{21} \ll \gamma_{21}$), the SQD is in the ground state, $\rho_{11} \approx 1$, while the other states are unpopulated, $\rho_{22} \approx \rho_{33} \approx 0$ [see Figs.~\ref{Fig2}(a)-(c)].
At higher magnitudes of $\Omega^0_{21}$ (more specifically, $\Omega^0_{21} \approx \gamma_{21}$), the $|1\rangle \rightarrow |2\rangle$ transition starts to develop, accompanied by a growing one-exciton population $\rho_{22}$ [Fig.~\ref{Fig2}(b)] and, respectively, depleting $\rho_{11}$ [Fig.~\ref{Fig2}(a)]. Upon increasing $\Omega^0_{21}$ further to $\Omega^0_{21} \gg \gamma_{21}$, the $|1\rangle \rightarrow |2\rangle$ transition saturates, which is reflected by plateaus in Figs.~\ref{Fig2}(a) and (b). However, in this regime, the population of the bi-exciton state remains close to zero, because the applied field magnitude is still smaller than the bi-exciton   binding energy, $\Omega^0_{21} < \Delta_B$ (for the parameters chosen, $\Delta_B = 10^{2.92}~\gamma_{21}$).
For $\Omega^0_{21} \gg \Delta_B$, both transitions are saturated, which explains the further reduction of $\rho_{11}$ and $\rho_{22}$.

The tiny peak in the field dependence of the $\rho_{11}$ population in the vicinity of $\Omega^0_{21} \approx \Delta_B/2 = 10^{2.62} \gamma_{21}$, i.e., at the frequency of the coherent two-photon transition, demands extra attention. Under this condition, the population of the ground state is directly transferred to the bi-exciton state, without populating the single exciton state. This breaks the saturation balance between the ground and single exciton states and leads to a further redistribution of populations, giving rise to the tiny peak.

The field dependence of the magnitude of the SQD's dipole moment $|p_{SQD}| = |\mu_{21}\rho_{21} + \mu_{32}\rho_{32}|$ [Fig.~\ref{Fig2}(d)] can be fully inferred from the field behaviour of the populations. As is observed, $|p_{SQD}|$ shows two peaks. For the external field magnitude $\Omega^0_{21} \ll \gamma_{21}$, the SQD is predominantly in its ground state $|1 \rangle$, whereas the populations of the single-exciton $|2 \rangle$ and bi-exciton   $|3 \rangle$ states are minor. Correspondingly, the coherences $\rho_{21}$ and $\rho_{32}$ are close to zero. When the transition $|1\rangle \leftrightarrow |2\rangle$ starts to develop, the coherence $\rho_{21}$ increases and reaches its maximum at $\Omega^0_{21} \approx \gamma_{21}$. Upon further increasing $\Omega^0_{21}$, the transition $|1 \rangle \rightarrow |2 \rangle$ becomes saturated, resulting in drop in $\rho_{21}$, as the latter is proportional to $\rho_{22} - \rho_{11}$. For the external field amplitude $\gamma_{21} \ll \Omega^0_{21} \ll \Delta_B$, the coherence $\rho_{32}$ is still negligible. This explains the low-field peak of $|p_{SQD}|$; it is mainly due to the field dependence of $\rho_{21}$.

In the strong saturation regime, when $\Omega^0_{21} \sim \Delta_B$,
the transition between the single-exciton $|2 \rangle$ and bi-exciton $|3 \rangle$ states starts to grow efficiently, giving rise to the second peak of $|p_{SQD}|$. Importantly, the coherence $\rho_{21}$ is also not zero now and contributes again to the low-field peak of $|p_{SQD}|$. This explains the difference in amplitudes of low- and high-field peaks. At $\Omega^0_{21} \gg \Delta_B$, all transitions are saturated and both coherences, $\rho_{21}$ and $\rho_{32}$ decrease, resulting in $|p_{SQD}| \to 0$. Note that we discussed here the route of an adiabatic increase of the external field magnitude $\Omega^0_{21}$. In the opposite case, the system follows the reversed route, in full correspondence with the behaviour of the populations.

Figure~\ref{Fig3} shows the field dependent response of the system for $\omega = \omega_3/2 = \omega_2 - \Delta_B/2$, i.e., when the applied field is in resonance with the coherent two-photon transition. Some features of the response, which are sharply contrasted with the previous case, should be mentioned. First, the field dependence of all populations is almost monotonous, where $\rho_{11}$ and $\rho_{22}$ do not show a plateau [like in Fig.~\ref{Fig2}(a)] or a peak [like in Fig.~\ref{Fig2}(b)], respectively. Second, unlike the case of direct excitation of the one-exciton resonance, $\omega = \omega_2$ (Fig.~\ref{Fig2}), considerable changes in the response occur at the field magnitude $\Omega^{0}_{21} \approx 10\gamma_{21}$. It should also be noticed that, in spite of the fact that this magnitude is much smaller than $\Delta_B/2$, so that there is no real transition to the one-exciton state, this state experiences considerable changes together with the ground and bi-exciton states. We attribute this to the fact that excitation of the bi-exciton state is immediately followed by spontaneous transition to the one-exciton state, implying that they develop synchronously. The applied field amplitude $\Omega^0_{21}$ at which the populations start to grow significantly, can be estimated from equating the Rabi frequency and the rate of excitation for the coherent two-photon transition, both calculated in second order of the perturbation theory. For the energy level scheme of our model, this gives $\Omega^0_{21} \sim (\gamma_{32}\Delta_B)^{1/2} \approx 10 \gamma_{32}$ (see, e.g., Ref.~\cite{MalyshevPRA1998}), which is in good agreement with the numerical value (compare with Fig.~\ref{Fig3}). Recall that $\gamma_{32} \approx \gamma_{21}$.

Furthermore, at $\Omega^0_{21} \approx \Delta_B/2 = 10^{2.62}~\gamma_{12}$, the population of the one-exciton state $\rho_{22}$ slightly increases, accompanied by a decrease of the bi-exciton   state population $\rho_{33}$. This happens because $\Omega^0_{21}$ overcomes the detuning away from the one-exciton resonance, which gives rise to the ladder-like population transfer between states $|1\rangle$, $|2\rangle$, and $|3\rangle$ and to the respective redistribution of their populations.

As for the case of $\omega = \omega_2$, the magnitude of the SQD's dipole moment, $|p_{SQD}|$, shows two peaks [Fig.~\ref{Fig3}(d)] with different amplitudes. Also in this case, these peaks originate from the creation of the coherences $\rho_{21}$ and $\rho_{32}$.
It should be noticed, however, that at the current frequency and for $\Omega^0_{21} \ll \Delta_B$, this occurs in two steps, because these coherences can not be created directly: the excitation is out of resonance with the $|1 \rangle \rightarrow |2 \rangle$ and $|2 \rangle \rightarrow |3 \rangle$ transitions. The incoming field first excites the state $|3 \rangle$, which is followed by the consecutive spontaneous transitions $|3 \rangle \rightarrow |2 \rangle$ and $|2 \rangle \rightarrow |1 \rangle$. Simultaneously, the coherences $\rho_{21}$ and $\rho_{32}$ are building up [see Eqs.~(\ref{all_dR1})]. Upon further increasing the applied field magnitude to $\Omega^0_{21} \approx \Delta_B/2 \approx 10^{2.62}~\gamma_{12}$ \, ($\hbar\Omega^{0}_{21} \approx 1.25~\mathrm{meV}$), the ladder transitions to the states $|2\rangle$ and $|3 \rangle$ become available, producing again both $\rho_{21}$ and $\rho_{32}$. As a consequence, the second peak of $|p_{SQD}|$ appears. The peak positions perfectly follow changes in the populations.
\begin{figure}[ht!]
\begin{center}
\includegraphics[width=0.8\columnwidth]{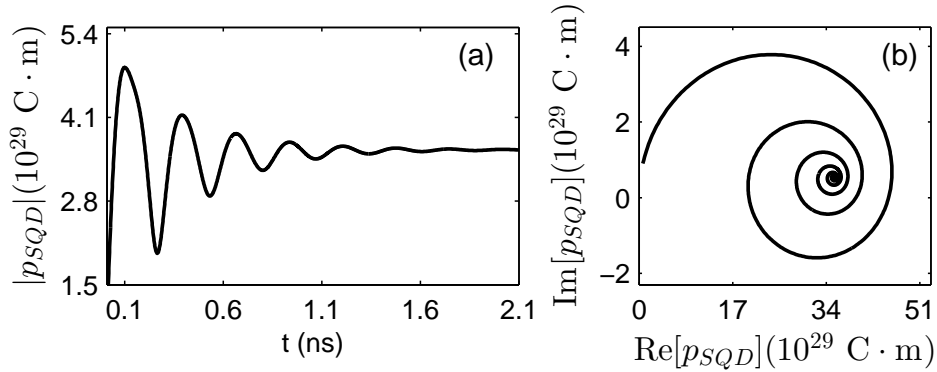}
\end{center}
\caption{Dynamics of the isolated SQD calculated for the excitation at $\omega = \omega_3/2$ with the incident field amplitude $\Omega^{0}_{21} = 10^{0.5}\gamma_{21}$. (a) Time evolution of the dipole moment magnitude $|p_{SQD}|$.
(b) Phase map of $p_{SQD}$ in the ($\mathrm{Re}[p_{SQD}], \mathrm{Im}[p_{SQD}]$ plane.
Other parameters are the same as in Fig.~\ref{Fig2}.
}
\label{Fig4}
\end{figure}

To further demonstrate the stability of the system, we calculated the kinetics of $|p_{SQD}|$ for $\Omega^{0}_{21} = 10^{0.5}\gamma_{21}$ and $\omega = \omega_2$. Figure~\ref{Fig4} shows the results. As is seen from panel (a), oscillations of $|p_{SQD}|$ are damped and reach their steady state after $t = 2~\mathrm{ns}$, i.e., approximately after $\gamma_{21}^{-1}$. The frequency of oscillations is on the order of $\Omega^{0}_{21}$. The phase space map [Fig.~\ref{Fig4}(b)] confirms the above kinetic picture, showing that the phase trajectory spirals to an equilibrium point.

\subsection{SQD-MNP hybrid}
\label{Hybrid SQD}
\begin{figure}[ht]
\begin{center}
\includegraphics[width=0.8\columnwidth]{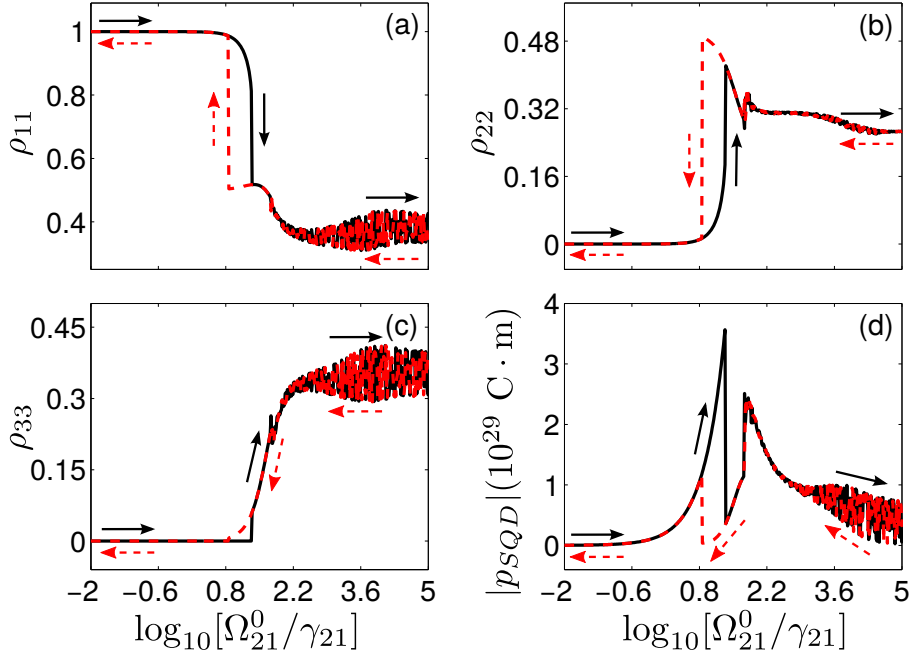}
\end{center}
\caption{Field dependent response of a strongly coupled SQD and MNP, center-to-center distance $d=16~\mathrm{nm}$. The applied field is in resonance with the bare $|1\rangle \rightarrow |2\rangle$ transition, $\omega=\omega_2$. Populations of (a) the ground state $\rho_{11}$, (b) single-exciton state $\rho_{22}$, and (c) bi-exciton  state $\rho_{33}$. (d) SQD's dipole moment magnitude, $|p_{SQD}|$. Solid and dashed arrows show the direction of sweeping the applied field magnitude $\Omega^0_{21}$ up and down, respectively.
}
\label{Fig5}
\end{figure}

\begin{figure}[ht!]
\begin{center}
\includegraphics[width=0.8\columnwidth]{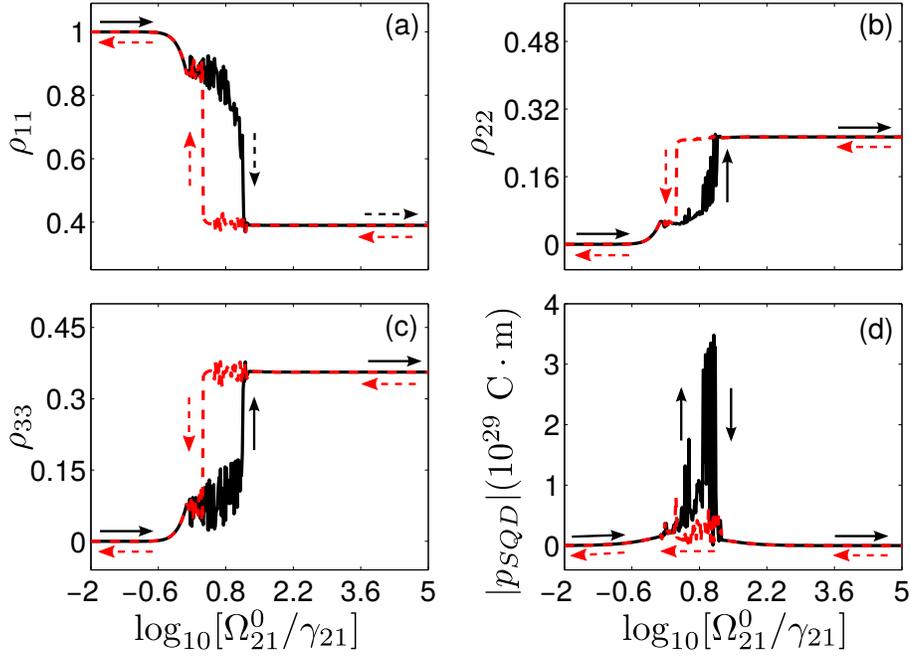}
\end{center}
\caption{Same as in Fig.~\ref{Fig5}, but now for the applied field tuned in resonance with the resonant two-photon transition, $\omega = \omega_3/2$.
}
\label{Fig6}
\end{figure}

Hereafter, we consider the case of a coupled SQD-MNP system by reducing the center-to-center distance to $d = 16~\mathrm{nm}$. Other parameters are the same as in Sec.~\ref{Isolated SQD}. For this value of $d$, the feedback parameters are found to be
$\hbar G_1 = (10^{2.72} + 10^{2.70}~i)\hbar\gamma_{21} = (1.6 + 1.5~i)~\mathrm{meV}$, $\hbar G_2 = (10^{2.84} + 10^{2.82}~i)~\hbar\gamma_{21} = (2.1 + 2.0~i)~\mathrm{meV}$, and $\hbar G_3 = (10^{2.78} + 10^{2.76}~i)~\hbar\gamma_{21} = (1.8 + 1.7~i)~\mathrm{meV}$.
We performed calculations similar to those presented in Sec.~\ref{Isolated SQD}, i.e., sweeping slowly the input field magnitude $\Omega^{0}_{21}$ up and down for two resonance conditions: $\omega = \omega_2$ and $\omega=\omega_3/2$. The results are presented in Figs.~\ref{Fig5} and ~\ref{Fig6}, respectively. As can be seen, the hybrid's optical response in both cases markedly differs from the response of an isolated SQD (compare with Fig.~\ref{Fig2} and Fig.~\ref{Fig3}).

We start with an analysis of the hybrid's response, when the applied field is in resonance with the one-exciton transition, $\omega=\omega_2$ (see Fig.~\ref{Fig5}). In contrast with an isolated SQD (compare with Fig.~\ref{Fig2}), the hybrid exhibits
two remarkable features. First, within a certain range of the input field amplitude $\Omega^{0}_{21} \ll \Delta_B$, its response shows hysteresis characteristic of bistability. In this range of $\Omega^{0}_{21}$, the population of the bi-exciton   state $\rho_{33}$ remains close to zero, as, due to the condition $\Omega^{0}_{21} \ll \Delta_B$, the transition efficiency to this state is negligible. Thus, the SQD to a good approximation then can be considered as a two-level system. The origin of bistability for a two-level emitter strongly coupled to an MNP has been discussed in a number of papers (see, e.g., Refs.~\cite{MalyshevPRB2011,MalyshevPRA2012,NugrohoJCP2013,ArtusoNL2008}): the strong SQD-MNP coupling is responsible for the occurrence of this effect. On a quantitative basis, it is driven by the feedback parameter $G_1$. The real and imaginary parts of $G_1$ (describing different mechanisms of bistability) should meet certain threshold conditions,~\cite{MalyshevPRB2011,MalyshevPRA2012,NugrohoJCP2013} which are fulfilled in the current case.

At higher $\Omega^{0}_{21}$, when the transition to the bi-exciton   state is well developed, the hybrid's response shows an unusual behaviour: the field dependence of the populations $\rho_{11}$, $\rho_{22}$, $\rho_{33}$, and the dipole moment magnitude $|p_{SQD}|$ appear as a "thick" noisy line (compare with Fig.~\ref{Fig2}), indicating the occurrence of instabilities in the hybrid's response. When sweeping down the field amplitude $\Omega^{0}_{21}$, the system follows another path, showing "bi-instability".
\begin{figure}[ht!]
\begin{center}
\includegraphics[width=0.8\columnwidth]{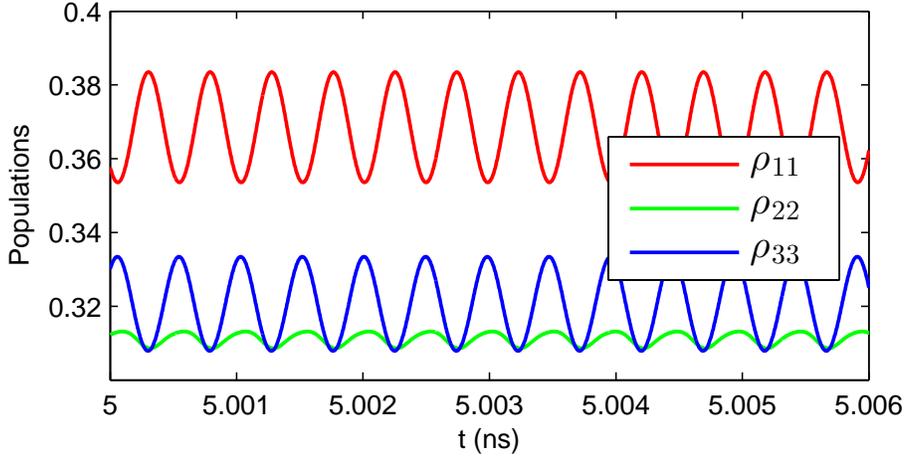}
\end{center}
\caption{Time evolution of the populations $\rho_{11}$, $\rho_{22}$, and $\rho_{33}$ after transient effects have disappeared, calculated for the applied field amplitude $\Omega_{21}=10^{2.2}\gamma_{21}$. Other parameters are the same as in Fig.~\ref{Fig5}.}
\label{Fig7}
\end{figure}

For excitation into the coherent two-photon resonance, $\omega = \omega_3/2$ (see Fig.~\ref{Fig6}), the hybrid's optical response shows similar features as in the previous case, $\omega = \omega_2$. However, details of the system's behaviour differ substantially. First of all, bistability and instabilities coexist within the same range of the applied field amplitude $\Omega^0_{21}$, i.e., both branches of the hysteresis loop turn out to be unstable, although not to the extent that the hysteresis disappears.
Second, the saturation of the system occurs also just within this range [see panels (a), (b), and (c) in Fig.~\ref{Fig6}]. Remember, that in the case of excitation into the one-exciton resonance ($\omega = \omega_2$), the SQD saturates at much larger applied field amplitudes $\Omega^0_{21}$ [see Fig.~\ref{Fig5}(a)-(c)]. However, at $\omega = \omega_3/2$, the hybrid is stable here. We see an explanation of these peculiarities in a higher efficiency of the coherent two-photon transition as compared to the ladder-like transitions at $\omega = \omega_3/2$, where both (real) transitions $|1\rangle \rightarrow |2\rangle$ and $|2\rangle \rightarrow |3\rangle$ are out of resonance with the applied field.

\begin{figure}[ht!]
\begin{center}
\includegraphics[width=0.8\columnwidth]{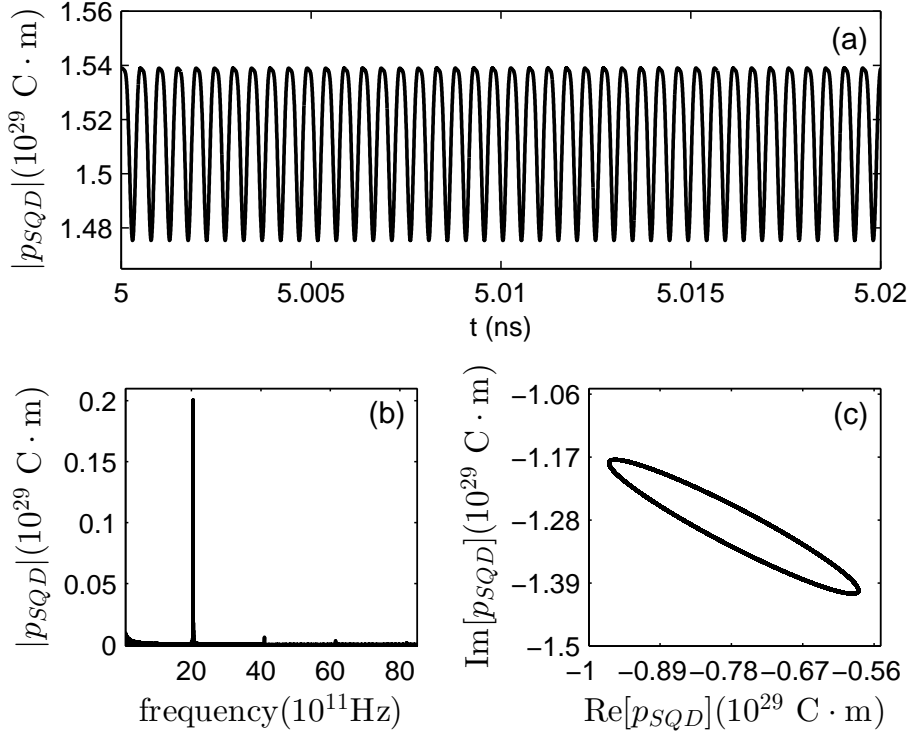}
\end{center}
\caption{(a) Time evolution of the magnitude of the SQD's dipole moment $|p_{SQD}|$, (b) Fourier spectrum of $|p_{SQD}|$, (c) phase space map of $p_{SQD}$ plotted in the ($\mathrm{Re}[p_{SQD}]$, $\mathrm{Im}[p_{SQD}]$) plane. The applied field amplitude is given by $\Omega^0_{21} = 10^{2.2} \gamma_{21}$. Other parameters are the same as in Fig.~\ref{Fig5}.}
\label{Fig8}
\end{figure}

To get insight into the nature of these instabilities, we calculated $\rho_{11}$, $\rho_{22}$, $\rho_{33}$, and $|p_{SQD}|$ at a particular amplitude of the applied field $\Omega^{0}_{21}$, lying within the range of interest. The results obtained for $\Omega^{0}_{21} = 10^{2.2}\gamma_{21}$ are shown in Fig.~\ref{Fig7} and Fig.~\ref{Fig8}. As is seen from Fig.~\ref{Fig7}, all populations exhibit sustained oscillations (self-oscillations), in spite of the fact that the input field amplitude $\Omega^{0}_{21}$ is fixed. We do not present in Fig.~\ref{Fig7} the transient stage of the population dynamics, before the oscillation regime sets in.  Note that $\rho_{11}$ and $\rho_{33}$ oscillate in anti-phase and have amplitudes significantly larger than that for $\rho_{22}$ oscillations. This suggests that efficient population transfer occurs between the ground and bi-exciton   states via the two-photon absorption, presumably, the ladder-like one (see below).

More information about the self-oscillation regime is provided by studying the magnitude of the SQD's dipole moment $p_{SQD}$. In Fig.~\ref{Fig8}, we plotted the time evolution of $|p_{SQD}|$ [panel (a)], its the Fourier spectrum [panel (b)], and the phase map of $p_{SQD}$ [panel (c)]. Consistent with the dynamics of the populations, $|p_{SQD}|$ shows sustained oscillations too, Fig.~\ref{Fig8} (a). In the Fourier spectrum of $|p_{SQD}|$, several harmonics are visible, however, one low-frequency is dominant; just this harmonic is responsible for the self-oscillation regime. Accordingly, the phase space map of $p_{SQD}$ represents an ellipse, because $\mathrm{Re} [p_{SQD}] \neq \mathrm{Im} [p_{SQD}]$. Note that the period of the phase space cycle for the parameters used is on the order of hundreds of femtoseconds. In other words, the SQD-MNP heterodimer in the self-oscillation regime generates a train of ultrashort optical pulses.

We also performed calculations for higher applied field magnitudes $\Omega^0_{21}$ to check out whether the self-oscillation regime is the unique regime in the high-field limit $\Omega^0_{21} > \Delta_B$. The results (not presented here) confirmed this. We only found that the frequency of self-oscillations depends on $\Omega^0_{21}$.

While the bistable regime of the hybrid can be analysed and understood in detail (see, e.g., Refs.~\cite{NugrohoJCP2013}), the physical picture of self-oscillations is much more complicated, because all three SQD's states are involved in this process. The complexity comes from the fact that both allowed transitions undergo renormalization due to the exciton-plasmon interaction: for the lower one, $\omega_{2} \rightarrow \omega_{2} + G_1^R Z_{21}$ and $\gamma_{21}/2 \rightarrow \gamma_{21}/2 - G_1^I Z_{21}$, and for the higher one, $\omega_{3} \rightarrow \omega_{3} + G_2^R Z_{32}$ and $(\gamma_{21} + \gamma_{32})/2 \rightarrow (\gamma_{21} + \gamma_{32})/2 - G_2^I Z_{32}$. So for $G_1^R, G_2^R, G_1^I, G_2^I > 0$ (our case), $\omega_{2}$ and $\omega_{3}$ acquire a red-shift and additional broadening, depending on the population differences $Z_{21}$ and $Z_{32}$, respectively. In the low-field limit ($\Omega^0_{21} \ll \gamma_{21}$), the population difference $Z_{21} \approx -1$, whereas $Z_{32} \approx 0$, because the states $|2 \rangle$ and $|3 \rangle$ are unpopulated. Accordingly, only the lower ($|1 \rangle \leftrightarrow |2 \rangle$) transition experiences the above mentioned renormalization:
$\omega_{2} \rightarrow \omega_{2} - G_1^R$ and $\gamma_{21}/2 \rightarrow \gamma_{21}/2 + G_1^I$, whereas the higher ($|2 \rangle \leftrightarrow |3 \rangle$) transition is kept unchanged.

After the SQD has switched from the lower to the upper branch (both stable) of the bistable characteristics (see Fig.~\ref{Fig5}), the population difference $Z_{21}$ becomes approximately zero, because the transition $|1 \rangle \leftrightarrow |2 \rangle$ is saturated~\cite{MalyshevPRB2011,NugrohoJCP2013}. Consequently, the renormalized transition frequency  $\omega_{2} + G_1^R Z_{21}$ and the relaxation rate  $\gamma_{21}/2 - G_1^I Z_{21}$ take their bar values $\omega_2$ and $\gamma_{21}/2$, respectively. At the same time, the state $|3 \rangle$ persists to be unpopulated [see. Fig.~\ref{Fig5} (c)]. However, as now $\rho_{22} \approx 1/2$, the $|2 \rangle \leftrightarrow |3 \rangle$ transition appears to be renormalized: $\omega_{3} \rightarrow \omega_{3} - (1/2)G_2^R$ and $(\gamma_{21} + \gamma_{32})/2 \rightarrow (\gamma_{21} + \gamma_{32})/2 + (1/2) G_2^I$, implying that the transition acquires a red shift $(1/2) G_2^R$ and an additional relaxation rate $(1/2) G_2^I$, both on the order of one meV, i.e. the bi-excitonic shift $\Delta_B$. Furthermore, when the amplitude of the external field exceeds the value of $\Delta_B$, which happens at approximately $\Omega^0_{21} = 10^{2.2}\gamma_{21}$, the one-exiton-to-bi-exciton transition comes into play, and the bi-exciton state begins to get populated. Note that this happens at a lower applied field amplitude as compared to the isolated SQD, where the corresponding value of $\Omega^0_{21} = 10^{2.62}\gamma_{21}$ (see Fig.~\ref{Fig2}). This effect originates from the plasmon-induced enhancement of the field acting on the SQD, as described by Eq.~(\ref{Omega21}).

The excitation of the bi-exciton state leads to a redistribution of populations of all states and, consequently, to changes in the population differences $Z_{21}$ and $Z_{32}$, giving rise to changing dynamic energy shifts and relaxation rates of the transitions. These dynamic modulations of the SQD parameters provoke a dynamic competition between two transitions, resulting finally in a self-oscillating response. Understanding these regimes requires special methods of the theory of nonlinear dynamical systems, such as Poincare maps (see, e.g., Ref.~\cite{Katok1997}), the application of which is beyond the scope of the present paper.

In our previous paper~\cite{NugrohoPRB2015}, we considered a model where the bi-excitonic shift is much larger than the dynamic shift and broadening of the quantum dot transitions induced by the interaction with the MNP. As a result, the latter are relevant only for the one-exciton transition. This leads to bistability in the one-exciton transition, but does not affect the two-exciton transition in a principle way (in Ref.~\cite{NugrohoPRB2015}, the only effect of the MNP on the two-exciton transition is its enhanced saturation). In the current paper, the bi-excitonic shift is on the order of the dynamic shift and broadening. As a result, a complicated interplay between the one-exciton and bi-exciton transitions occurs when the driving field starts to populate the bi-exciton state, resulting in the new phenomena of self-oscillations and ({\it vide infra}) chaos.

\begin{figure}[ht!]
\begin{center}
\includegraphics[width=0.8\columnwidth]{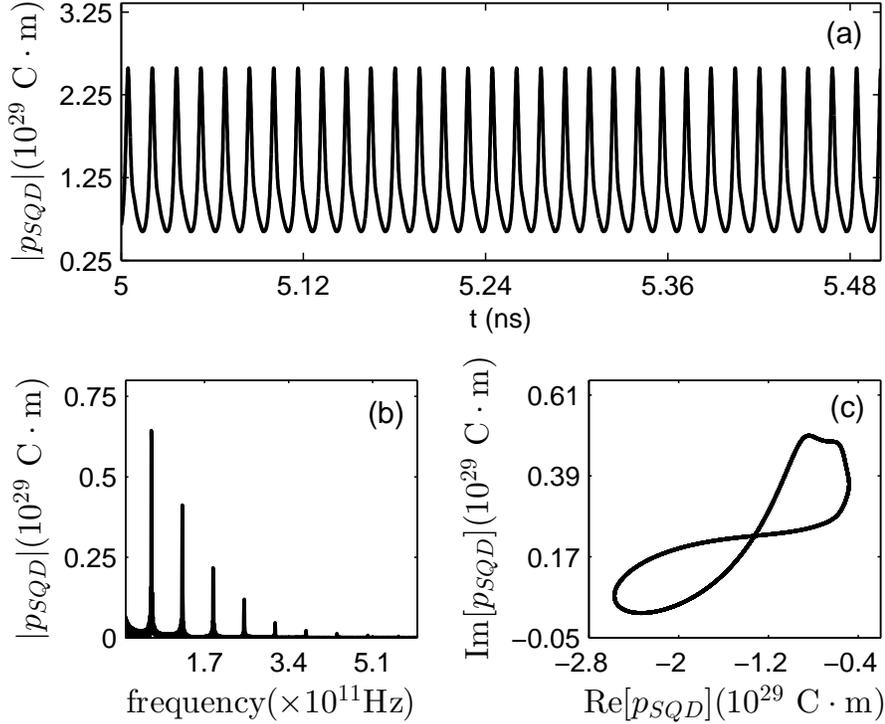}
\end{center}
\caption{(a) Time evolution of the magnitude of the SQD's dipole moment $|p_{SQD}|$, (b) Fourier spectrum of $|p_{SQD}|$, (c) phase space map of $p_{SQD}$ plotted in the ($\mathrm{Re}[p_{SQD}]$, $\mathrm{Im}[p_{SQD}]$) plane. In all plots we used the applied field amplitude $\Omega^0_{21} = 10^{0.8} \gamma_{21}$. Other parameters are the same as in Fig.~\ref{Fig6}.
}
\label{Fig9}
\end{figure}

When the applied field is in resonance with the coherent two-photon transition, $\omega = \omega_3/2 = \omega_2 - \Delta_B/2$ (see Fig.~\ref{Fig6}), the hybrid shows a response which is, in general, similar to the previous case in the sense that it also demonstrates a hysteresis loop at moderate applied field amplitudes $\Omega^0_{21}$, around a value of $\Omega^0_{21} \approx 10^{0.8}\gamma_{21}$, accompanied by a saturation regime for higher $\Omega^0_{21}$. However, in contrast to the case of excitation of the one-exciton resonance $(\omega = \omega_2)$, here instabilities occur within the hysteresis loop, showing however a stable response at values of $\Omega^0_{21}$ for which all transitions are saturated.
Note that the saturation regime starts at smaller amplitudes of $\Omega^0_{21}$ than for the case of an isolated SQD (compare with Fig.~\ref{Fig3}), which again is a consequence of the effect of the plasmon-induced enhancement of the applied field, Eq.~(\ref{Omega21}).

To get insight into the nature of the instabilities of the SQD response, we calculated the time evolution of the magnitude $|p_{SQD}|$ of the SQD dipole moment, its Fourier spectrum and phase space map at the lower branch of the hysteresis loop [see Fig.~\ref{Fig6}(d)] for $\Omega^0_{21} = 10^{0.8}\gamma_{21}$, which lies within the instability range. Figure~\ref{Fig9} shows the results. From plot (a), it is seen that $|p_{SQD}|$ exhibits sustained self-oscillations, but of more complicated form than for the case of excitation into the one-exciton resonance ($\omega = \omega_2$, see Fig.~\ref{Fig8}). The Fourier spectrum of $|p_{SQD}|$ [plot (b)] consists of a number of harmonics with comparable intensities. So the resulting time-dependent signal represents a superposition of these harmonics. The more complicated behaviour of $|p_{SQD}|$ is also reflected in the phase map of $p_{SQD}$: the trajectory of the system in the phase space [$\mathrm{Re}(p_{SQD}),\mathrm{Im}(p_{SQD})$] is now represented by a close $\infty$-like curve.
Calculations performed for other values of $\Omega^0_{21}$ revealed that the self-oscillation regime of the hybrid's response is the only type of instability that occurs in addition to the already known bistability.

Particular attention should be paid to the period of the self-oscillations. First, it depends on the applied field magnitude $\Omega^0_{21}$ and therefore can be controlled by changing $\Omega^0_{21}$. Second, the values of the period, for the parameters used, are on the order of several picoseconds at $\Omega^0_{21} = 10^{0.8} \gamma_{21}$  (see Fig.~\ref{Fig9}) end even significantly shorter (hundreds of femtoseconds) at $\Omega^0_{21} = 10^{2.2} \gamma_{21}$  (see Fig.~\ref{Fig8}). In other words, the heterodimer in the self-oscillation regime generates a train of ultrashort optical pulses, that might be interesting from the viewpoint of practical applications in nanodevices.

Summarizing our findings, we find that the feedback parameters $G_m$ play a crucial role in the optical response of the heterodimer, eventually resulting in instabilities in the optical response: bistability and self-oscillations. Note that for the set of parameters we used in our calculations, the real $G_m^R$ and imaginary $G_m^I$ parts of each $G_m$ are approximately equal to each other. The general question arises: what happens if they are considerably different? To answer this, we performed model calculations when $G_m^I = 0$, keeping $G_m^R$ the same as previously. This eliminates the dynamic damping described by the imaginary parts of $G_m$. We  will only discuss the results for the case when the applied field is in resonance with the coherent two-photon transition, $\omega = \omega_3/2$, just to illustrate the heterodimer's optical response in this particular situation.

\begin{figure}[ht!]
\begin{center}
\includegraphics[width=0.8\columnwidth]{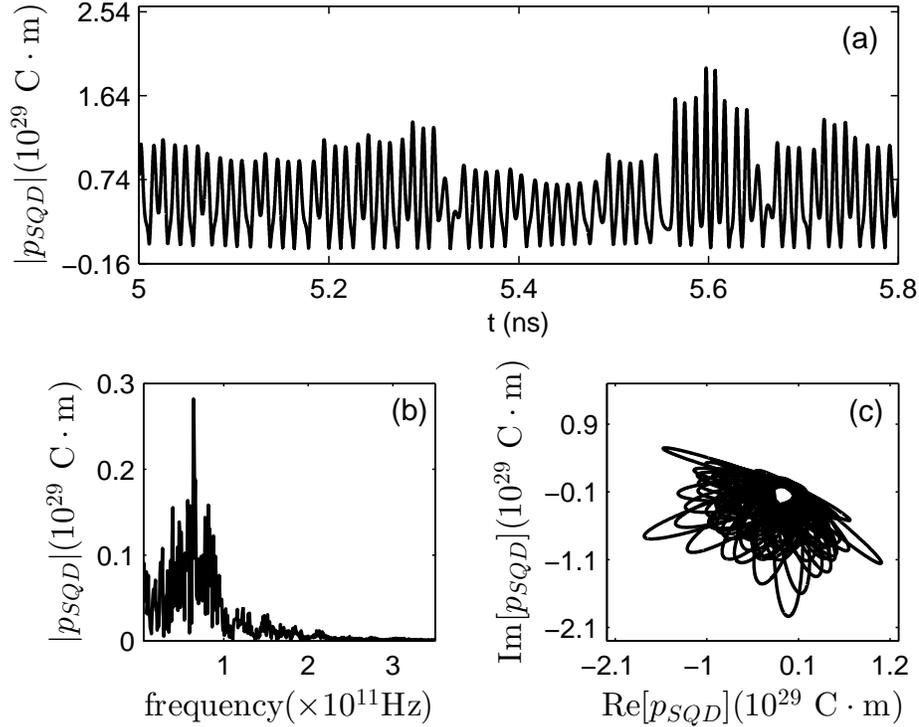}
\end{center}
\caption{(a) Time evolution of the SQD's dipole moment magnitude $|p_{SQD}|$, (b) Fourier spectrum of $|p_{SQD}|$, (c) phase space map of $p_{SQD}$ plotted in the ($\mathrm{Re}[p_{SQD}]$, $\mathrm{Im}[p_{SQD}]$) plane. The applied field is in resonance with the coherent two-photon transition, $\omega = \omega_3/2 = \omega_2 - \Delta_B/2$. Its amplitude $\Omega^0_{21} = 10^{-0.55} \gamma_{21}$. Other parameters are the same as in Fig.~\ref{Fig6}, except $G_m^I = 0$.
}
\label{Fig10}
\end{figure}

Without going into detail about the field dependence of the populations $\rho_{11}$, $\rho_{22}$, $\rho_{33}$, and the magnitude $|p_{SQD}|$ of the  SQD dipole moment, we only present the dynamics of the dipole moment $|p_{SQD}|$, calculated for the applied field amplitude $\Omega^0_{21} = 10^{-0.55}\gamma_{21}$, see Fig.~\ref{Fig10}. From plot (a) we observe that the time evolution of $|p_{SQD}|$ reveals an oscillatory regime, the nature of which can be deduced from the Fourier spectrum [plot (b)] and the phase-space map of $p_{SQD}$ [plot (c)]. A is seen, the Fourier spectrum represents a broad structureless quasi-continuum, suggesting that the oscillatory regime we found is of quasi-chaotic nature. The phase-space map confirms this, showing a trajectory that covers a finite volume of the $p_{SQD}$ phase space, not forming any closed loop.

In the opposite case of zero real parts of $G_m$ (not shown here), the system exhibits only the self-oscillation regime. From this observation, we can conclude that the imaginary parts of the feedback parameters somehow stabilize the chaotic behaviour of the hybrid, turning this regime into one characterized by self-oscillations.

\section{Summary}
\label{summary}
We conducted a theoretical study of the optical response of a heterodimer comprising a semiconductor quantum dot and a metal nanosphere. In contrast to many preceding studies, where quantum dots have been treated as a two-level system with the ground state $|1\rangle$ and the one-exciton state $|2\rangle$, we also accounted for a bi-exciton (two coupled excitons) state $|3\rangle$. As was shown previously, the presence of a metal nanoparticle in close proximity to a (two-level) semiconductor quantum dot may result in optical bistability and hysteresis of the composite, driven by a complex-valued feedback parameter. This feature also persists for a heterodimer with a three-level quantum dot. However, including the third level gives rise to a much richer optical dynamics of the system, including self-oscillations and quasi-chaotic behaviour. The dynamics is governed by two complex-valued feedback parameters, associated with the ground-to-one exciton and one-exciton-to-bi-exciton   transitions. The real and imaginary parts of these feedback parameters determine different scenarios for the composite's optical instabilities. It should be stressed that all these effects occur under {\it single}-frequency continuous wave excitation. By contrast, the optical response of an isolated three-level quantum dot does not show any instability. We have presented the physical explanation of the instabilities, which find their origin in the competition between the ground-to-one exciton and one exciton-to-bi-exciton  transitions, driven by the quantum dot self-action via the metal nanoparticle. We performed our calculations for a model system that may be realized in practice: a heterodimer comprised of a closely spaced ZnS/ZnSe core-shell quantum dot and a silver nanosphere and found bistability and self-oscillations in the optical response. Another candidate to observe the instabilities would be a heterodimer comprised of an $\mathrm{In_x}\mathrm{Ga_{1-x}}$As/GaAs quantum dot and a triangular silver nanoparticle, absorbing in a wide spectral range, from the visible to the infrared~\cite{WuNRL2015}.

To conclude, we note that the period of the self-oscillations depends on the applied field magnitude and can be pushed down to the sub-picosecond regime. Thus, potentially an SQD-MNP heterodimer, in its unstable regime, represents a tunable nanogenerator of trains of ultrashort optical pulses that might be interesting from the viewpoint of practical applications in nanodevices.

\ack
This work has been supported by NanoNextNL, a Micro- and Nano-technology consortium of the Government of the Netherlands and 130 partners.

\section*{References}
\bibliography{BibFile}

\providecommand{\newblock}{}
\begin{thebibliography}{10}
\expandafter\ifx\csname url\endcsname\relax
  \def\url#1{{\tt #1}}\fi
\expandafter\ifx\csname urlprefix\endcsname\relax\def\urlprefix{URL }\fi
\providecommand{\eprint}[2][]{\url{#2}}

\bibitem{FrankenPRL1961}
Franken P, Hill A, Peters C and Weinreich G 1961 {\em Phys. Rev. Lett.\/} {\bf
  7} 118

\bibitem{MaimanNature1960}
Maiman T~H 1960 {\em Nature\/} {\bf 187} 493

\bibitem{NieAM1993}
Nie W 1993 {\em Adv. Mater.\/} {\bf 5} 520

\bibitem{TominagaJCP1996}
Tominaga K and Yoshihara K 1996 {\em J. Chem. Phys.\/} {\bf 104} 4419

\bibitem{OronPRL2002}
Oron D, Dudovich N and Silberberg Y 2002 {\em Phys. Rev. Lett.\/} {\bf 89}
  273001

\bibitem{CotterScience1999}
Cotter D, Manning R, Blow K, Ellis A, Kelly A, Nesset D, Phillips I, Poustie A
  and Rogers D 1999 {\em Science\/} {\bf 286} 1523

\bibitem{NoskovNJP2012}
Noskov R, Krasnok A and Kivshar Y~S 2012 {\em New J. Phys.\/} {\bf 14} 093005

\bibitem{Boyd2003}
Boyd R~W 2003 {\em Nonlinear optics\/} (Academic press)

\bibitem{MaierAM2001}
Maier S~A, Brongersma M~L, Kik P~G, Meltzer S, Requicha A~A and Atwater H~A
  2001 {\em Adv. Mater.\/} {\bf 13} 1501

\bibitem{NovotnyNP2011}
Novotny L and Van~Hulst N 2011 {\em Nature Photon.\/} {\bf 5} 83--90

\bibitem{WangJPCB2006}
Wang X, Du Y, Ding S, Wang Q, Xiong G, Xie M, Shen X and Pang D 2006 {\em J.
  Phys. Chem. B\/} {\bf 110} 1566

\bibitem{GongJCP2006}
Gong H~M, Wang X~H, Du Y~M and Wang Q~Q 2006 {\em J. Chem. Phys.\/} {\bf 125}
  024707

\bibitem{FuAPL2012}
Fu M, Wang K, Long H, Yang G, Lu P, Hetsch F, Susha A~S and Rogach A~L 2012
  {\em Appl. Phys. Lett.\/} {\bf 100} 063117

\bibitem{KauranenNP2012}
Kauranen M and Zayats A~V 2012 {\em Nature Photon.\/} {\bf 6} 737

\bibitem{SinghNT2013}
Singh M~R 2013 {\em Nanotechnology\/} {\bf 24} 125701

\bibitem{ZhangPRL2006}
Zhang W, Govorov A~O and Bryant G~W 2006 {\em Phys. Rev. Lett.\/} {\bf 97}(14)
  146804

\bibitem{ArtusoNL2008}
Artuso R~D and Bryant G~W 2008 {\em Nano Lett.\/} {\bf 8} 2106

\bibitem{MalyshevPRB2011}
Malyshev A~V and Malyshev V~A 2011 {\em Phys. Rev. B\/} {\bf 84}(3) 035314

\bibitem{NugrohoJCP2013}
Nugroho B~S, Iskandar A~A, Malyshev V~A and Knoester J 2013 {\em J. Chem.
  Phys.\/} {\bf 139} 014303

\bibitem{LiOE2012}
Li J~B, Kim N~C, Cheng M~T, Zhou L, Hao Z~H and Wang Q~Q 2012 {\em Opt.
  Express\/} {\bf 20} 1856

\bibitem{RidolfoPRL2010}
Ridolfo A, Di~Stefano O, Fina N, Saija R and Savasta S 2010 {\em Phys. Rev.
  Lett.\/} {\bf 105}(26) 263601

\bibitem{ManjavacasNL2011}
Manjavacas A, Abajo F~J~G~d and Nordlander P 2011 {\em Nano Lett.\/} {\bf 11}
  2318

\bibitem{ZhangPRB2011}
Zhang W and Govorov A~O 2011 {\em Phys. Rev. B\/} {\bf 84} 081405

\bibitem{ArtusoPRB2013}
Artuso R~D and Bryant G~W 2013 {\em Phys. Rev. B\/} {\bf 87} 125423

\bibitem{CarrenoJAP2014}
Carre{\~n}o F, Ant{\'o}n M, Melle S, Calder{\'o}n O~G, Cabrera-Granado E, Cox
  J, Singh M~R and Egatz-G{\'o}mez A 2014 {\em J. Appl. Phys.\/} {\bf 115}
  064304

\bibitem{Ikeda1OC1979}
Ikeda K 1979 {\em Opt. Commun.\/} {\bf 30} 257

\bibitem{IkedaPRL1980}
Ikeda K, Daido H and Akimoto O 1980 {\em Phys. Rev. Lett.\/} {\bf 45} 709

\bibitem{GibbsPRL1981}
Gibbs H~M, Hopf F~A, Kaplan D~L and Shoemaker R~L 1981 {\em Phys. Rev. Lett.\/}
  {\bf 46} 474

\bibitem{SadeghiAPL2014}
Sadeghi S and Patty K 2014 {\em Appl. Phys. Lett.\/} {\bf 104} 083101

\bibitem{SadeghiNT2015}
Sadeghi S~M, Wing W~J and Gutha R~R 2015 {\em Nanotechnology\/} {\bf 26} 085202

\bibitem{SadeghiPRA2015}
Sadeghi S~M, Wing W~J and Gutha R~R 2015 {\em Phys. Rev. A\/} {\bf 92} 023808

\bibitem{SadeghiJNR2016}
Sadeghi S 2016 {\em J. Nanopart. Res.\/} {\bf 18} 1--9

\bibitem{NugrohoPRB2015}
Nugroho B~S, Malyshev V~A and Knoester J 2015 {\em Phys. Rev. B\/} {\bf 92}
  165432

\bibitem{bohren2008}
Bohren C~F and Huffman D~R 2008 {\em Absorption and scattering of light by
  small particles\/} (John Wiley \& Sons)

\bibitem{maier2007}
Maier S~A 2007 {\em Plasmonics: Fundamentals and Applications\/} (Springer)

\bibitem{SchollNat2012}
Scholl J~A, Koh A~L and Dionne J~A 2012 {\em Nature\/} {\bf 483} 421

\bibitem{StuflerPRB2006}
Stufler S, Machnikowski P, Ester P, Bichler M, Axt V~M, Kuhn T and Zrenner A
  2006 {\em Phys. Rev. B\/} {\bf 73} 125304

\bibitem{JundtPRL2008}
Jundt G, Robledo L, H{\"o}gele A, F{\"a}lt S and Imamo{\u{g}}lu A 2008 {\em
  Phys. Rev. Lett.\/} {\bf 100} 177401

\bibitem{GerardotNJP2009}
Gerardot B, Brunner D, Dalgarno P, Karrai K, Badolato A, Petroff P and
  Warburton R 2009 {\em New J. Phys.\/} {\bf 11} 013028

\bibitem{YanPRB2008}
Yan J~Y, Zhang W, Duan S, Zhao X~G and Govorov A~O 2008 {\em Phys. Rev. B\/}
  {\bf 77}(16) 165301

\bibitem{ArtusoPRB2011}
Artuso R~D, Bryant G~W, Garcia-Etxarri A and Aizpurua J 2011 {\em Phys. Rev.
  B\/} {\bf 83} 235406

\bibitem{JohnsonPRB1972}
Johnson P~B and Christy R~W 1972 {\em Phys. Rev. B\/} {\bf 6}(12) 4370

\bibitem{SenJPCS2010}
Sen P, Chattopadhyay S, Andrews J~T and Sen P~K 2010 {\em J. Phys. Chem.
  Solids\/} {\bf 71} 1201--1205

\bibitem{MalyshevPRA1998}
Malyshev V~A, Glaeske H and Feller K~H 1998 {\em Phys. Rev. A\/} {\bf 58} 670

\bibitem{MalyshevPRA2012}
Malyshev A~V 2012 {\em Phys. Rev. A\/} {\bf 86} 065804

\bibitem{Katok1997}
Katok A and Hasselblatt B 1997 {\em Introduction to the modern theory of
  dynamical systems\/} vol~54 (Cambridge university press)

\bibitem{WuNRL2015}
Wu C, Zhou X and Wei J 2015 {\em Nanoscale Res. Lett.\/} {\bf 10} 354

\end{thebibliography}

\end{document}